\documentclass[conference]{IEEEtran}
\usepackage{graphicx}
\usepackage{multirow}
\usepackage{amsmath}
\usepackage{subfigure}

\providecommand{\e}[1]{\ensuremath{\times 10^{#1}}}

\begin{document}
\title{New Approximate Multiplier for Low Power\\ Digital Signal Processing}

\author{\IEEEauthorblockN{Farzad Farshchi, Muhammad Saeed Abrishami, and Sied Mehdi Fakhraie}
\IEEEauthorblockA{School of Electrical and Computer Engineering\\
University of Tehran\\
Tehran, Iran\\
\{f.farshchi, msabrishami, fakhraie\}@ut.ac.ir}}



\maketitle

\begin{abstract}
%\boldmath
In this paper a low power multiplier is proposed. The proposed multiplier utilizes Broken-Array Multiplier approximation method on the conventional modified Booth multiplier. This method reduces the total power consumption of multiplier up to 58\% at the cost of a small decrease in output accuracy. The proposed multiplier is compared with other approximate multipliers in terms of power consumption and accuracy. Furthermore, to have a better evaluation of the proposed multiplier efficiency, it has been used in designing a 30-tap low-pass FIR filter and the power consumption and accuracy are compared with that of a filter with conventional booth multipliers. The simulation results show a 17.1\% power reduction at the cost of only 0.4dB decrease in the output SNR.\\
\end{abstract}


\begin{IEEEkeywords}
Approximate computimg, low power, DSP systems, FIR filter, inaccurate hardware units
\end{IEEEkeywords}

% \fi
\IEEEpeerreviewmaketitle

%%%%%%%%%%%%%%%%%%%%%%%%%%%%%%%%%%%
%					Introduction
%%%%%%%%%%%%%%%%%%%%%%%%%%%%%%%%%%%
\section{Introduction}
Power consumption is one of the most important characteristics of any electronic device especially for battery powered hand-held devices. A lot of efforts have been put into reducing power consumption of systems at different design levels.

One of the favorite techniques for power reduction is trading accuracy for power consumption. Different designs have been proposed in this regard. One of these approaches is using approximate computing in applications showing inherent error resilience. Some DSP, multimedia, fuzzy logic, neural networks, wireless communications, recognition, and data mining algorithms are examples of such applications \cite{mahdiani}, \cite{venka}.

Approximation can be performed using different techniques such as allowing some timing violations (e.g., voltage over-scaling or over-clocking) and function approximation techniques (e.g., modifying the Boolean function of a circuit) or a mixture \cite{venka}. Reference \cite{mahdiani} proposed an approximate adder and an approximate multiplier based on a technique named Broken-Array Multiplier (BAM) and demonstrated their benefits in terms of delay and area when exploited to implement a face recognition neural network and defuzzification block of a fuzzy processor. In \cite{kulk}, another approximate multiplier was proposed. It consisted of some $2\times2$ inaccurate building blocks and could save power between 31.8\% and 45.4\% over an accurate multiplier. The proposed multiplier was used to filter an image. The approximate filter saved power by 41.5\% over an accurate one and achieved a Signal to Noise Ratio (SNR) of 20.4dB.  Reference \cite{kelly} designed an approximate signed 32-bit multiplier for speculation purposes in pipelined processors. The multiplier is 20\% faster, with a probability of error around 14\%. In \cite{yin},  Error Tolerant Multiplier (ETM) was introduced. It computed the approximate result by dividing multiplication into one accurate and one approximate part.  Accuracy for various bit-width multipliers was reported.  Power saving of more than 50\% was reported for a 12-bit multiplier. The authors did not demonstrate any application for their design.  The authors in \cite{ning} proposed the Error Tolerant Adder (ETA) by dividing the operation into precise and approximate parts and proposed a new circuit for the approximate part. They improved Power-Delay Product (PDP) more than 65\% comparing to conventional adders. An FFT processor was implemented with ETA to compare the quality reduction of the output and results showed the output quality reduction.  A quantitative criterion on the quality loss and power saving of the whole system was not reported. Three forms of approximate Full Adders (FAs) were introduced in \cite{gupta}. These FAs were used to build adders of a DCT-IDCT processor in image compression applications. The proposed approximate blocks improved the power consumption of the system by about 50\% while at the same time leaded to about 6dB Peak Signal to Noise Ratio (PSNR) reduction. In \cite{sampson}, a Java extension for a compiler was proposed to map some parts of a code to approximate hardware, so that less power is consumed. An approximate hardware architecture was also introduced. The authors in \cite{moha} introduced meta-functions that behave gracefully under voltage over scaling. These meta-functions construct the main parts of some multimedia, recognition, and data mining algorithms. Reference \cite{venka}  proposed a methodology for modeling and analysis of circuits for approximate computing. This method can be used to analyze how an approximate circuit behaves with reference to an accurate implementation.

Most of the previous work utilized inefficient arithmetic units for applying their approximation techniques which obviously leads to inefficient approximate arithmetic units.  In addition, multipliers have the greatest share of arithmetic unit power consumption in most DSP systems, but many of the previous works focused on the less power consuming units like adders neglecting the total system power reduction. In this paper, an approximate modified Booth multiplier is proposed. The approximation technique is based on BAM \cite{mahdiani}. Due to better efficiency of modified Booth multiplier comparing to other multipliers, it is expected that the approximate version is also more efficient comparing to other approximate multipliers. To examine the proposed multiplier, it is utilized in design of a low-pass 30-tap FIR filter. The results of implementation are compared with filters built out of accurate multipliers with different Word Lengths (WLs).

Rest of this paper is organized as follows. In Section II, the proposed Broken-Booth Multiplier is introduced and the output error is analyzed. Section III shows the synthesis and simulation results and compares the results with other multipliers. Section IV concludes this paper.

%%%%%%%%%%%%%%%%%%%%%%%%%%%%%%%%%%%
%					Broken-Booth Multiplier
%%%%%%%%%%%%%%%%%%%%%%%%%%%%%%%%%%%
\section{Broken-Booth Multiplier}
In this section we introduce the approximation algorithm and discuss the statistical parameters of the output error. The evaluation method of the proposed multiplier is also discussed in this section.

\subsection{Approximation Algorithm}
Fig. 1 shows the dot diagram notation of the proposed approximate modified Booth signed multiplier \cite{weste}. Every row demonstrates one of the Partial Products (PPs) of the multiplication. In this approximation method, all the dot products positioned at the right hand side of the Vertical Breaking Level (VBL) are replaced by zero. In modified Booth algorithm, 2's complements of some of the PPs are required. This means, complementing the PP then adding one to it. In this figure, 'S' will be one if the 2's complement is required and otherwise it will be zero. According to this, two breaking algorithms are possible. Fig. 1 (a) shows the first possible method. In this method, which we will call it Broken-Booth Multiplier Type0 through the rest of this paper, the required PPs are complemented and added with one then breaking procedure is applied. Fig. 1 (b) shows the second possible method. We will call this method Broken-Booth Multiplier Type1 through the rest of this paper. In this method the required PPs are complemented but are not added with one at this stage. The breaking procedure is applied after this stage and then the result is added with one if that one is not replaced by zero during the breakage. In both methods after this stage the PPs are added according to their positions. Complementing a row needs an increment operation, therefore nullifying some sign bits Type1 results in less increment operations, thus more power saving. The weakness of this method is higher inaccuracy penalty in comparison with Type0.

% Figure fig_bbm
\begin{figure}[!t]
\centering
\includegraphics[width=3in]{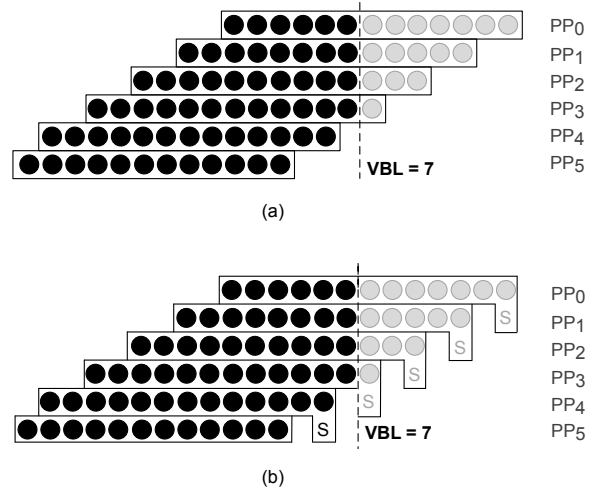}
\caption{The Broken-Booth Multiplier Type0 (a) and Type1 (b) for $\mathit{WL}=12$ and $\mathit{VBL}=7$. Adopted from \cite{weste}.}
\label{fig_bbm}
\end{figure}

\subsection{Statistical Parameters of the Output Error}
The statistical parameters of the output error of the Broken-Booth Multiplier Type0 with WL of 12, are reflected in Table I for different VBLs. Throughout this paper we use the term “error” instead of “output error”.  In this table, the error is calculated according to Eq. (1) and Mean Squared Error (MSE) to Eq. (2).

% Eq 1
\begin{equation}
error=approximate \: output-accurate \: output.
\label{eq1}
\end{equation}

% Table tbl_err
\begin{table}[!t]
  \centering
  \renewcommand{\arraystretch}{1.3}
  \caption{MSE, Error Mean and Probability and Minimum Error of the Broken-Booth Multiplier Type0 with WL = 12.}
  \label{tbl_err}
    \begin{tabular}{|c|c|c|c|c|}
    \hline
          & Error Mean & MSE  & Error Prob. & Min-Error \\
    \hline
    VBL = 3 & -3.50  & 2.22\e{1}  & 0.6875 & -1.10\e{1} \\
    \hline
    VBL = 6 & -6.15\e{1} & 5.05\e{3} & 0.9375 & -1.71\e{2} \\
    \hline
    VBL = 9 & -7.89\e{2} & 7.52\e{5} & 0.9893 & -2.22\e{3} \\
    \hline
    VBL = 12 & -8.53\e{3} & 8.33\e{7} & 0.9983 & -2.32\e{4} \\
    \hline
    \end{tabular}
\end{table}

To obtain these parameters, the arithmetic behavior of the multiplier is modeled and in a simulation environment, all the possible input vectors are exhaustively applied to it. For example, error percentage distribution of the Broken-Booth multiplier with WL = 10 and VBL = 9 is shown in Fig. 2. It should be noticed that in this figure the error is normalized to $2^{19}$ which is the maximum possible output of a $10\times10$ signed multiplier.

% Figure fig_err_hist
\begin{figure}[!t]
\centering
\includegraphics[width=3.5in]{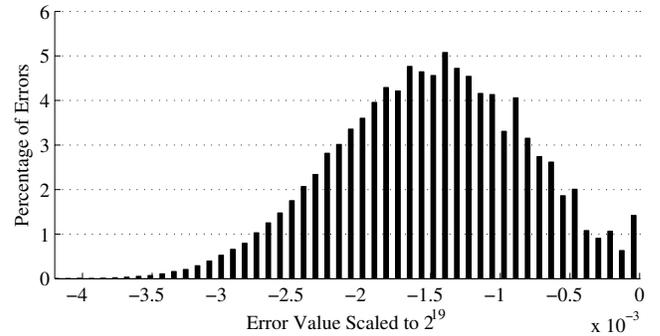}
\caption{The percentage of error distribution of Broken-Booth Multiplier Type0 for $\mathit{WL}=10$ and $\mathit{VBL}=9$.}
\label{fig_err_hist}
\end{figure}

In \cite{oppen}, in order to analyze the quantization error induced on the output of a DSP system, an analytic method is described. In this method the quantization error is assumed to be a white noise and as a result a power level is defined for it. We have evaluated the output error of our proposed multiplier and compare it to the previous work, based on this suggestion. Therefore, the most important parameter reported in Table I is MSE which is calculated using Eq (2).

% Eq 2
\begin{equation}
\mathit{MSE}=\frac{1}{N}\sum_{i=0}^{N-1}error^2(i).
\label{eq2}
\end{equation}

In this equation, i is the representative of input vector number and $N$ is the number of applied input vectors which is equal to $2^{24}$ for a $12\times12$ multiplier.

As seen in Table I, all the error parameters increase proportional to VBL. A similar trend exists for other word lengths.

\subsection{Evaluation Method}
To evaluate the proposed multiplier and also make an analogy between the previous designs and the proposed one, the hardware related parameters such as delay, area, and power consumption should be extracted. To do that, a parametric Verilog description of the design is developed. In this model, setting the VBL to 0 will result in an accurate version of the multiplier. Moreover, PPs are generated based on modified Booth algorithm and the summation of them is described at high level description and the details of implementation are left to synthesis tool. The design is synthesized in standard cells of 90nm CMOS technology using Synopsys Design Compiler. To calculate the power consumption of the synthesized circuit, the post-synthesis simulation is employed and a Value-Change-Dump (VCD) file is extracted. The extracted VCD file is fed into PrimeTime PX and the average total power - the sum of dynamic and leakage power - is reported.

%%%%%%%%%%%%%%%%%%%%%%%%%%%%%%%%%%%
%					Simulation Results
%%%%%%%%%%%%%%%%%%%%%%%%%%%%%%%%%%%
\section{Simulation Results}
In this section, first, a comparison between the hardware characteristics of the proposed multiplier and an accurate version is made. Next, the proposed multiplier is compared to the previous designs in the literature; finally, as an application an FIR filter is implemented once using an accurate multiplier and another time with the proposed approximate multiplier. To come up with an analogy, the output SNR and total power consumption of the filter for different cases are compared.

\subsection{Comparison with the Accurate Booth Multiplier}
At the first step, an accurate $16\times16$ Booth multiplier is obtained by setting the VBL to 0 in the developed Verilog model of Broken-Booth Multiplier. Next, the model is synthesized and the minimum possible delay (T$_{min}$) is obtained. After that, both the approximate (Type0) and accurate models are synthesized with timing constraints of T$_{min}$ and four different timing constraints more relaxed than T$_{min}$. It should be noticed that in the approximate model, the VBL parameter is set to 15, i.e. from 32 columns, 15 are nullified. Furthermore to compare the performance of the proposed multiplier and the accurate one, the proposed multiplier is synthesized again for minimum possible delay. The power consumption is calculated for both multipliers and reflected in Fig. 3 for each delay setting. The simulation is done on synthesized models of the multipliers and in this process the circuits are tested with 5\e{5} random input vectors. Since the input vectors are generated randomly, the rate of switching activity for internal nodes is relatively higher than applying a runtime workload. However, as the condition is the same for all models, the comparison between power consumptions remains valid.

% Figure fig_power-delay
\begin{figure}[!t]
\centering
\includegraphics[width=3.5in]{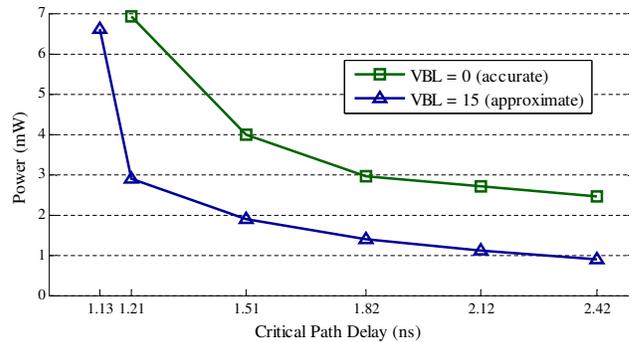}
\caption{Total power vs. delay for accurate and approximate multipliers with input $\mathit{WL}=16$.}
\label{fig_power-delay}
\end{figure}

It can be inferred from Fig. 3 that the power consumption of the Broken-Booth Multiplier is about half of the power consumption of the accurate one. The power consumption of both multipliers grows suddenly as the delay reaches its minimum value. The minimum possible delays for the accurate and Broken-Booth multipliers are 1.21ns and 1.13ns, respectively. Therefore, the Broken-Booth multiplier is 6.6\% faster than the accurate one.

The Broken-Booth Multiplier is also compared to the accurate one in this way for different WLs and Tables II and III demonstrate the percent of power and area reduction of the Broken-Booth Multiplier respectively, comparing to the accurate multiplier. As shown in Table II, the power consumption of the Broken-Booth Multiplier is reduced by 28.4\% to 58.6\% and the area is reduced by 19.7\% to 41.8\%. As the multipliers hardware almost halved on average, it is expected that the area and power consumption reduce by this rate. For example, in the case WL = 12 and VBL = 11, 36 bits out of 77 are nullified which results in removing some parts of PP generators and PP adders, therefore we expect that area and power consumption almost reduce by 47\% ($36\div77$). Moreover, as the power reduction is more than the area reduction and total capacitance is proportional to area, it is concluded that in Broken-Booth Multiplier the switching activities of the internal nodes are also reduced.
% Table tbl_power
\begin{table}[!t]
\renewcommand{\arraystretch}{1.1}
\setlength{\tabcolsep}{1pt}
\caption{Percentage of Power Reduction for Various WLs and Delay Constraints.}
\label{tbl_power}
\centering
\begin{tabular}{|c|c|c|c|c|c|c|}
\hline
& $1\!\times\!T_{min}$ & $1.25\!\times\!T_{min}$ & $1.5\!\times\!T_{min}$ & $1.75\!\times\!T_{min}$ & $2\!\times\!T_{min}$ & Mean\\
& (\%) & (\%) & (\%) & (\%) & (\%) & (\%)\\
\hline
\hline
WL=4, & \multirow{2}{*}{18.2} & \multirow{2}{*}{35.8} & \multirow{2}{*}{33.9} & \multirow{2}{*}{27.6} & \multirow{2}{*}{24.7} & \multirow{2}{*}{28.0}\\
VBL=3 & & & & & &\\
\hline
WL=8, & \multirow{2}{*}{44.8} & \multirow{2}{*}{47.7} & \multirow{2}{*}{58.0} & \multirow{2}{*}{64.2} & \multirow{2}{*}{66.9} & \multirow{2}{*}{56.3}\\
VBL=7 & & & & & &\\
\hline
WL=12, & \multirow{2}{*}{52.7} & \multirow{2}{*}{52.2} & \multirow{2}{*}{60.0} & \multirow{2}{*}{57.3} & \multirow{2}{*}{70.9} & \multirow{2}{*}{58.6}\\
VBL=11 & & & & & &\\
\hline
WL=16, & \multirow{2}{*}{58.1} & \multirow{2}{*}{52.8} & \multirow{2}{*}{53.2} & \multirow{2}{*}{59.0} & \multirow{2}{*}{64.0} & \multirow{2}{*}{57.4}\\
VBL=15 & & & & & &\\
\hline
\end{tabular}
\end{table}

% Table tbl_area
\begin{table}[!t]
\renewcommand{\arraystretch}{1.1}
\setlength{\tabcolsep}{1pt}
\caption{Percentage of Area Reduction for Various WLs and Delay Constraints.}
\label{tbl_area}
\centering
\begin{tabular}{|c|c|c|c|c|c|c|}
\hline
& $1\!\times\!T_{min}$ & $1.25\!\times\!T_{min}$ & $1.5\!\times\!T_{min}$ & $1.75\!\times\!T_{min}$ & $2\!\times\!T_{min}$ & Mean\\
& (\%) & (\%) & (\%) & (\%) & (\%) & (\%)\\
\hline
\hline
WL=4, & \multirow{2}{*}{14.0} & \multirow{2}{*}{25.0} & \multirow{2}{*}{19.0} & \multirow{2}{*}{21.6} & \multirow{2}{*}{18.9} & \multirow{2}{*}{19.7}\\
VBL=3 & & & & & &\\
\hline
WL=8, & \multirow{2}{*}{45.3} & \multirow{2}{*}{31.6} & \multirow{2}{*}{28.9} & \multirow{2}{*}{29.9} & \multirow{2}{*}{31.2} & \multirow{2}{*}{33.4}\\
VBL=7 & & & & & &\\
\hline
WL=12, & \multirow{2}{*}{54.0} & \multirow{2}{*}{41.8} & \multirow{2}{*}{39.9} & \multirow{2}{*}{33.3} & \multirow{2}{*}{40.0} & \multirow{2}{*}{41.8}\\
VBL=11 & & & & & &\\
\hline
WL=16, & \multirow{2}{*}{53.8} & \multirow{2}{*}{42.8} & \multirow{2}{*}{38.3} & \multirow{2}{*}{37.1} & \multirow{2}{*}{36.0} & \multirow{2}{*}{41.6}\\
VBL=15 & & & & & &\\
\hline
\end{tabular}
\end{table}

\subsection{Comparison to Previous Designs}
In order to compare the proposed multiplier with other approximate multipliers,  two approximate multipliers presented in \cite{mahdiani} and \cite{kulk}, are modeled, synthesized, and simulated in the same technology and compared to the proposed one in terms of Power-Delay Product (PDP) and MSE. The multiplier with lower PDP and error power is preferred. In \cite{mahdiani}, one of the proposed methods which is named Broken-Array Multiplier (BAM) is an unsigned approximate multiplier. In BAM, in addition to VBL, there is another parameter called Horizontal Breaking Level (HBL) for adjusting precision and hardware saving. In this comparison we set the HBL to 0 and only manipulate the VBL.  It should be noticed that there is no difference between BAM and its signed counterpart, in terms of MSE. The multiplier presented in \cite{kulk} is another unsigned approximate multiplier made up from basic blocks of $2\times2$ approximate multipliers. In \cite{kulk}, there is no defined parameter to adjust the precision of multiplier. Hence, in our implementation, we modified the design and defined the K parameter as illustrated in Fig. 4. In this method, an imaginary vertical line is introduced between the PPs, and the blocks positioned entirely on the right hand side of this line are replaced by approximate blocks and K controls the position of this line. In fact, the K parameter acts so alike to VBL in our proposed method and enhances the versatility of the design in \cite{kulk}. To compare the multipliers, the Verilog descriptions of both models are developed in a parametric manner.

% Figure fig_kulk
\begin{figure}[!t]
\centering
\includegraphics[width=3.3in]{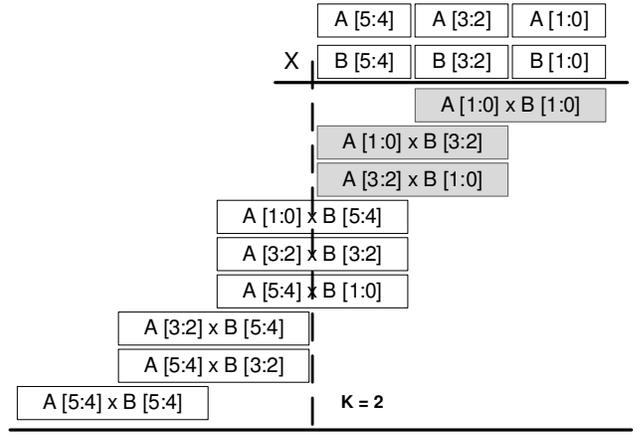}
\caption{The PP diagram of multiplier \cite{kulk} with our added parameter K for WL = 6. Blocks in gray are approximate.}
\label{fig_kulk}
\end{figure}

To calculate the PDP over MSE, the following procedure is taken:

\begin{enumerate}
\item Using the method introduced in Section II.A, the MSE of the multipliers is calculated over five different precision settings.
\item All multipliers with each precision setting are synthesized for minimum delay; the power consumption and PDP of each synthesis result is calculated afterwards.
\item The synthesis procedure is repeated once again with timing constraint of 1.75ns. In this step the PDP would be the product of calculated power consumption and 1.75ns.
\item The average PDP is calculated from the results of steps 2 and 3.
\end{enumerate}

%1)  Using the method introduced in Section II.A, the MSE of the multipliers is calculated over five different precision settings.
%
%2)  All multipliers with each precision setting are synthesized for minimum delay; the power consumption and PDP of each synthesis result is calculated afterwards.
%
%3)  The synthesis procedure is repeated once again with timing constraint of 1.75ns. In this step the PDP would be the product of calculated power consumption and 1.75ns.
%
%4)  The average PDP is calculated from the results of steps 2 and 3.

Fig. 5 shows the different PDPs as calculated in steps 2 through 4 over the MSE and adjusting parameter, corresponding to each of them. It shows that the variation of PDP over MSE is different for the steps 2 and 3.

% fig_comp
\begin{figure*}[!t]
\centering
\subfigure[]{\includegraphics[width=3.5in]{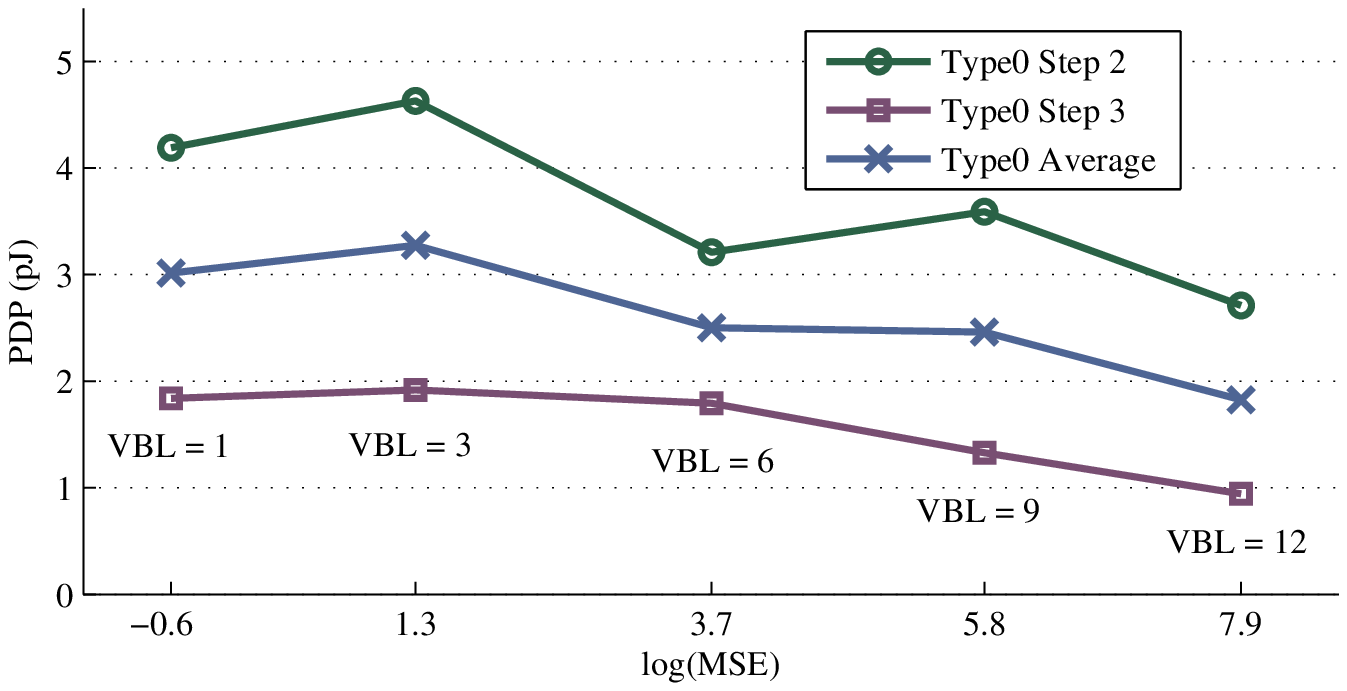}}
\hfil
\subfigure[]{\includegraphics[width=3.5in]{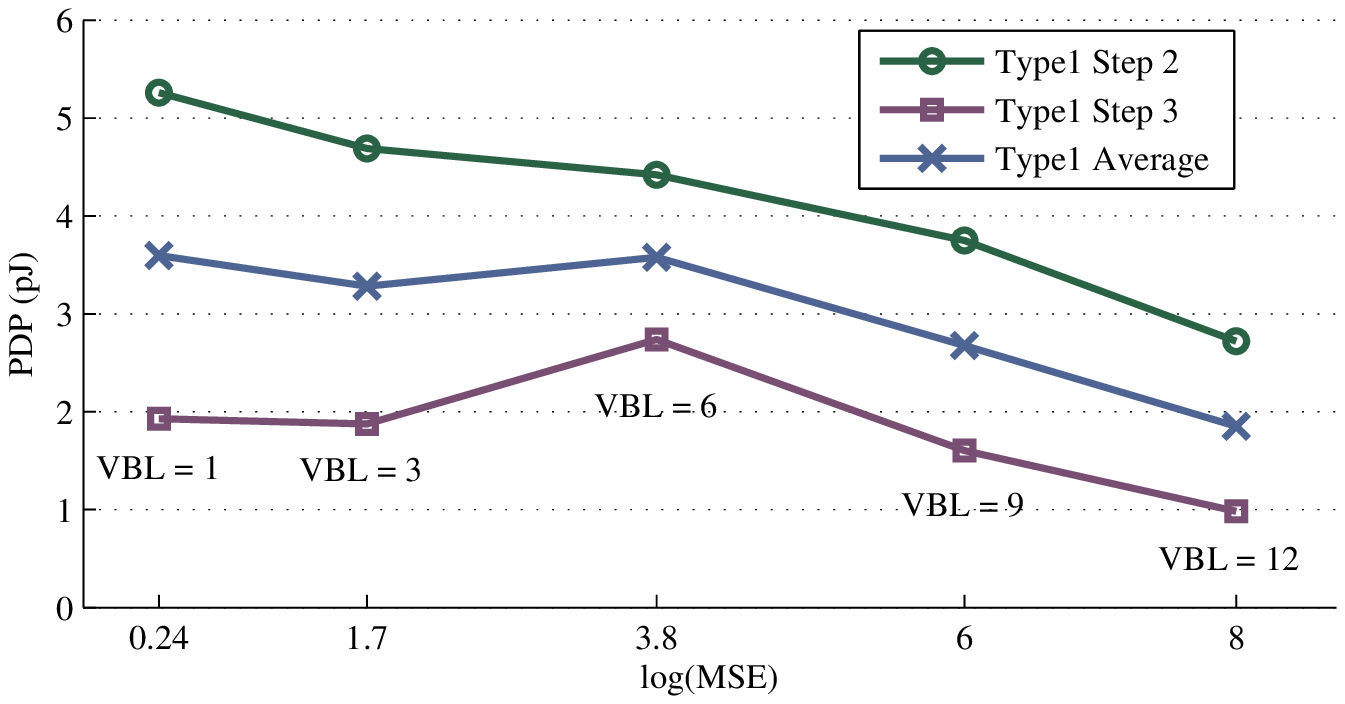}}
\vfil
\subfigure[]{\includegraphics[width=3.5in]{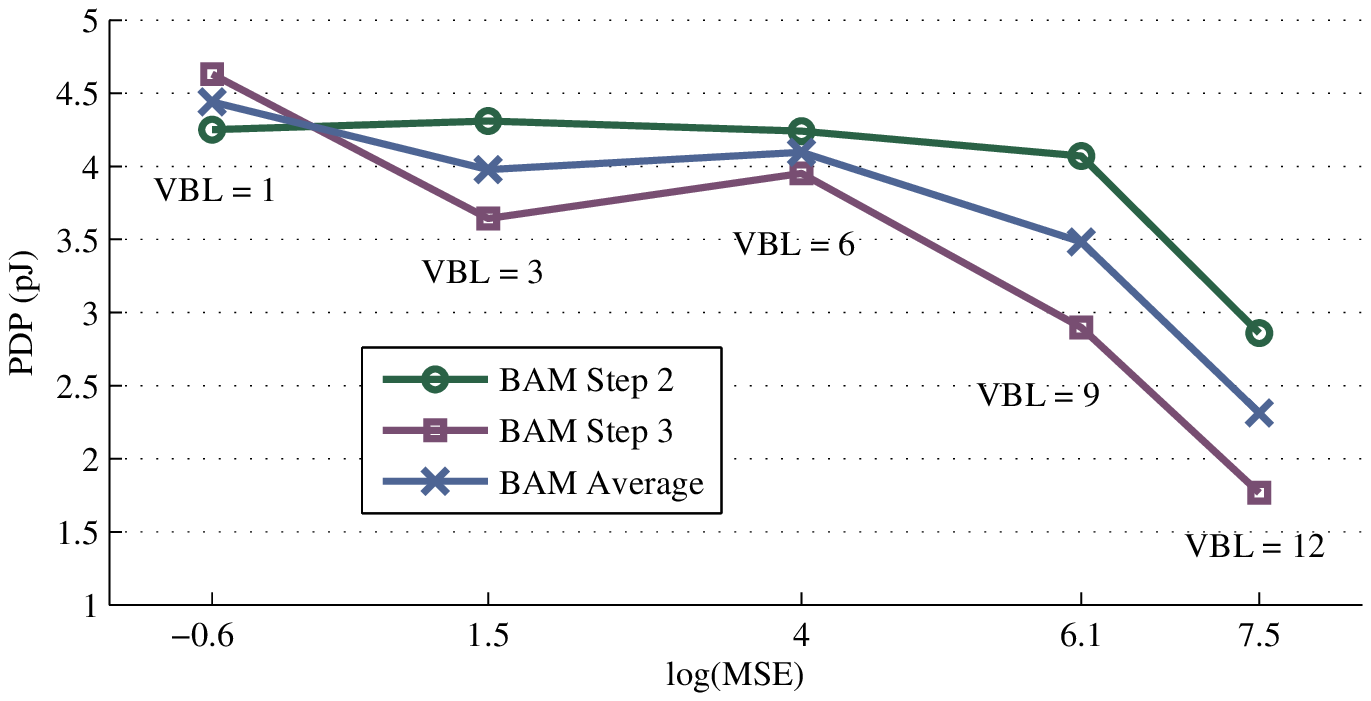}}
\hfil
\subfigure[]{\includegraphics[width=3.5in]{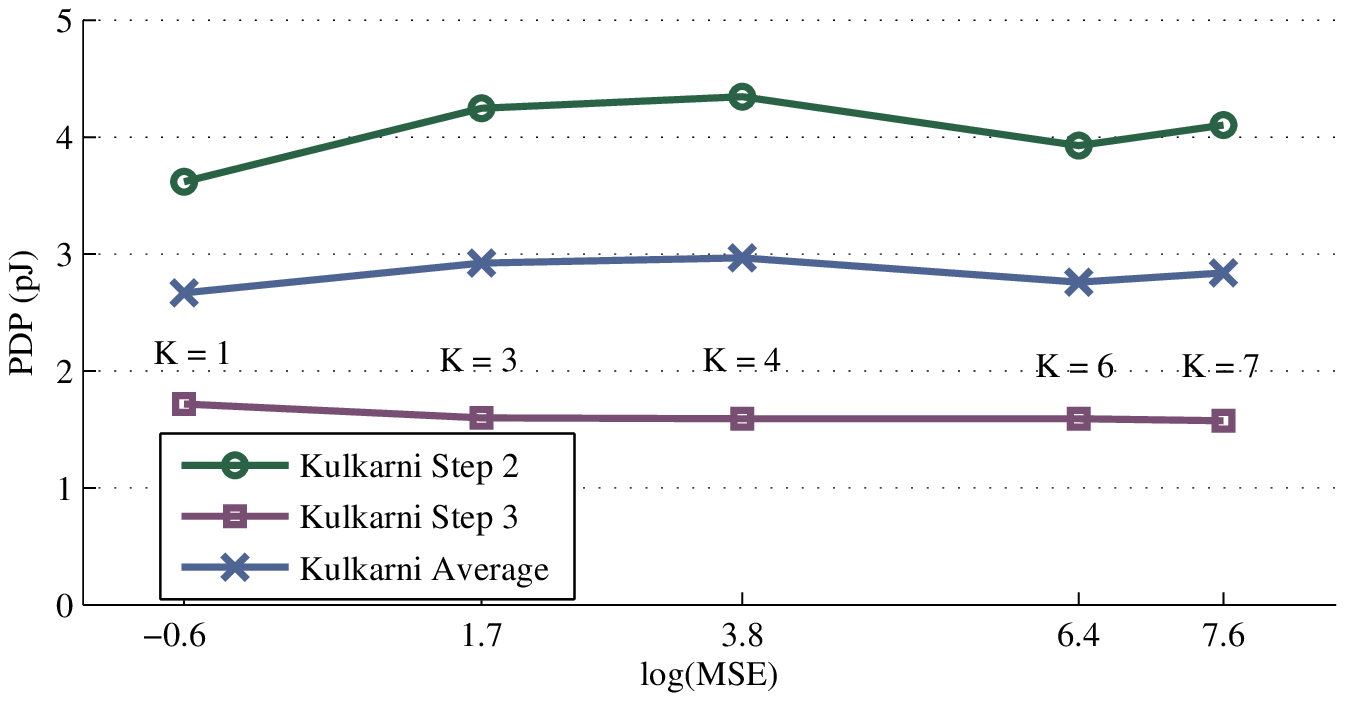}}
\caption{PDP vs. MSE logarithm for studied multipliers, Broken-Booth Multiplier Type0 (a), Type1 (b), BAM (c), and The multiplier in \cite{kulk} (d).}
\label{fig_comp}
\end{figure*}

In Fig. 6 the calculated average PDP of each multiplier is depicted over MSE in a single diagram. The multiplier in \cite{kulk} has the best PDP at lower MSE but as the error power increases, it does not show any PDP improvement. The Broken-Booth Multipliers Type0 and Type1 have better PDP for high MSE values comparing to \cite{kulk} and the PDP of them decreases almost steadily as the MSE grows. The reduction of PDP for Type0 is more graceful than Type1. The reason could be ability of synthesis tool to optimize the implied circuits in step 3.

% Figure fig_ave_all
\begin{figure}[!t]
\centering
\includegraphics[width=3.5in]{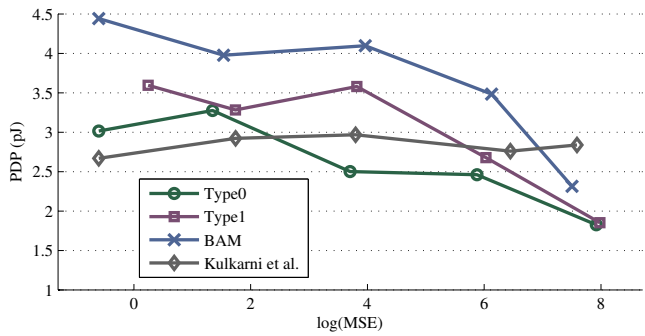}
\caption{Average PDP vs. MSE logarithm for studied multipliers.}
\label{fig_ave_all}
\end{figure}

\subsection{FIR Filter}
The application we have used is a low-pass FIR filter which is introduced in \cite{shim}. Fig. 7 shows the block diagram of the testbed, frequency response of the filter $H(\omega)$, and test signals $d_i (\omega)$. The testbed is designed based on using the filter in a real situation. The bandwidth and guard bandwidth of signals $d_i[n]$ are 0.25$\pi$  and 0.1$\pi$, respectively. The FIR filter input is the sum of signals $d_i[n]$ in the presence of white Gaussian noise source with -30dB power spectral density, $\eta[n]$. The $d_2[n]$  and $d_3[n]$ signals are located on transition and stop bands respectively. The desired signal is $d_1[n]$ which is located on pass band region. The SNR at the output and input of filter is defined as, SNR$_{out}=10\log_{10}\frac{\sigma^2_{d_1}}{\sigma^2_{d_1-y}}$ and
SNR$_{in}=10\log_{10}\frac{\sigma^2_{d_1}}{\sigma^2_{d_1-x}}$, respectively, where $\sigma^2_{d_1-y}=E[|d_1-y|^2]$ and $\sigma^2_{d_1-x}=E[|d_1-x|^2]$.

% Figure fig_shim
\begin{figure}[!t]
\centering
\subfigure[]{\includegraphics[width=3.5in]{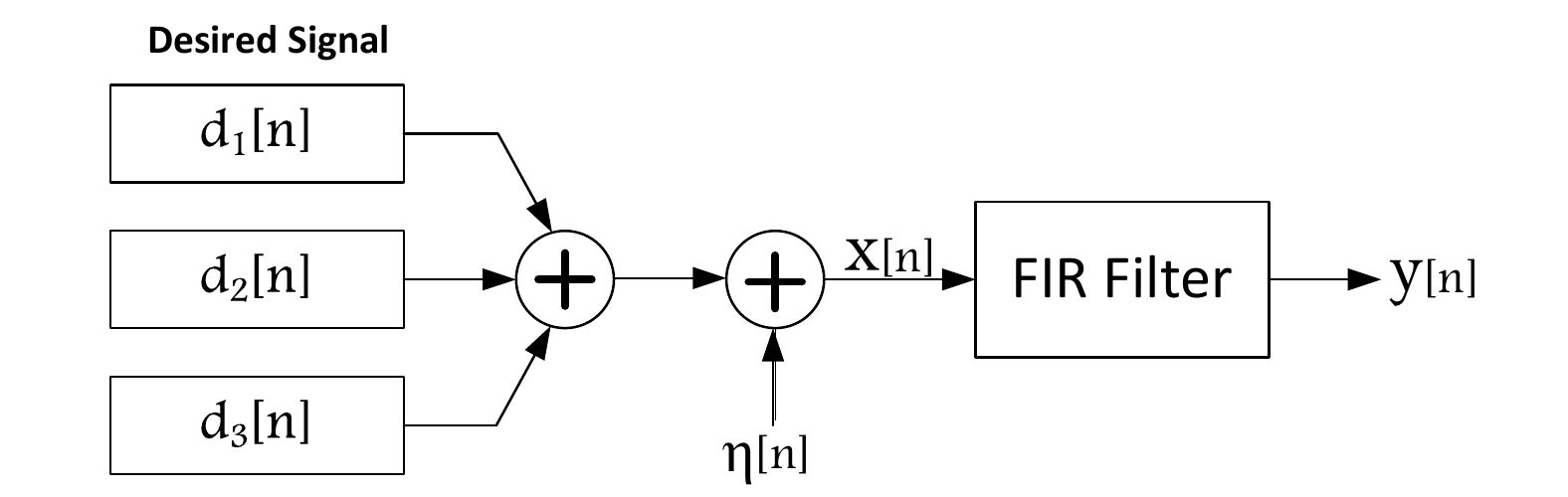}}
\vfill
\subfigure[]{\includegraphics[width=3.5in]{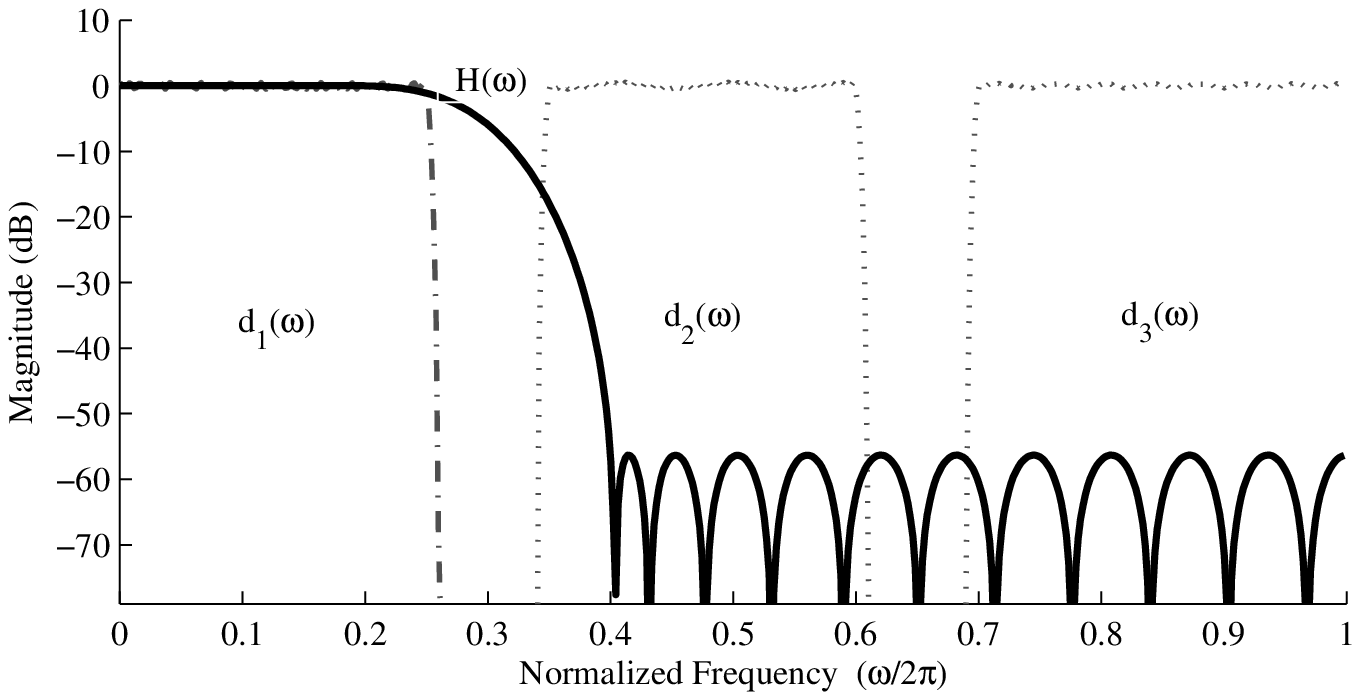}}
\caption{Testbed for simulating the FIR filter (a) frequency response of the filter and input signals (b) \cite{shim}.}
\label{fig_shim}
\end{figure}

A 30-tap order Parks-McClellan low-pass filter is modeled with double precision arithmetic. Simulation results show SNR$_{out}$ = 25.7dB and SNR$_{in}$ = -3.47dB. It means this filter increases SNR$_{out}$ up to 29.1dB in comparison with SNR$_{in}$. Next, a fixed-point filter is modeled with variable WL. Fig. 8 (a) shows SNR$_{out}$  for different WLs. Since Booth multipliers are optimum for even WLs, simulations are done based on even WLs. For the sake of efficient hardware implantation, the least possible WL should be chosen. Therefore, we choose WL = 16 and SNR$_{out}$ = 25.4dB, as is implied from Fig. 8 (a) that lower WLs lead to significant SNR$_{out}$ reduction.

% Figure fig_snr
\begin{figure}[!t]
\centering
\subfigure[]{\includegraphics[width=3.1in]{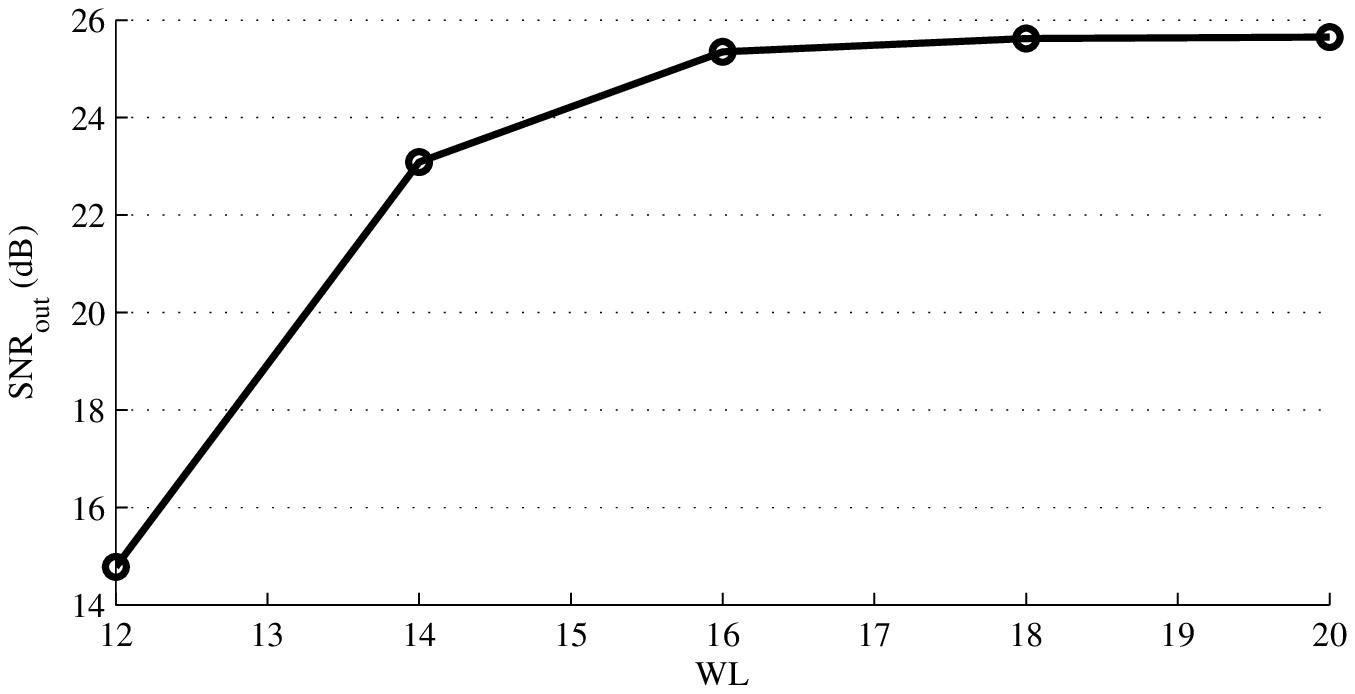}}
\vfill
\subfigure[]{\includegraphics[width=3.1in]{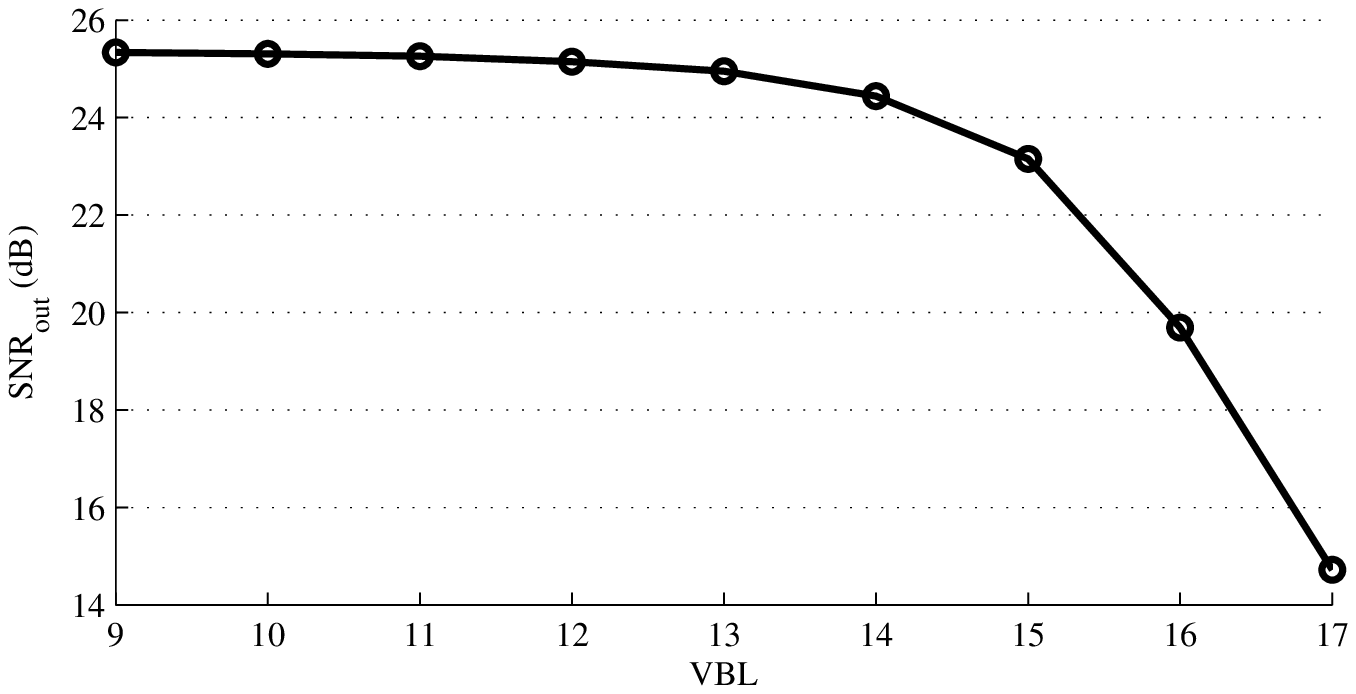}}
\caption{SNR$_{out}$ vs. WL (a)  SNR$_{out}$ vs. VBL (b).}
\label{fig_snr}
\end{figure}

Next, the Broken-Booth Multiplier Type0 is used as filter's multipliers. Fig. 8 (b) illustrates SNR$_{out}$ for different VBLs. It is seen that increasing VBL, leads to steady SNR$_{out}$ reduction. The desired operating point is defined by VBL = 13 and SNR$_{out}$ = 25dB for realization of the filter using Broken-Booth Multiplier, as higher VBL values leads to significant SNR$_{out}$ reduction. In order to demonstrate hardware reduction, the filter is modeled in Verilog with parametric WL and VBL. The model is synthesized for 3 cases:

\begin{enumerate}
\item WL = 16, VBL = 0,
\item WL = 16, VBL = 13,
\item WL = 14, VBL = 0.
\end{enumerate}

%1)  WL = 16, VBL = 0,
%
%2)  WL = 16, VBL = 13,
%
%3)  WL = 14, VBL = 0.

The case 3 is considered to evaluate further WL reduction effects on filter hardware and comparing with using Broken-Booth Multiplier at higher WL. Synthesis results, power consumption, and  SNR$_{out}$ are reported in Table IV. The value of QUAP in this table is introduced in \cite{gupta} and defined as:

% Eq 3
\begin{align}
&\mathit{QUAP} =\nonumber\\
&QUality \times Area\: savings\:(\%) \times Power\: savings\:(\%)
\label{eq3}
\end{align}

% Table tbl_fir_result
\begin{table}[!t]
\renewcommand{\arraystretch}{1.2}
\setlength{\tabcolsep}{2pt}
\caption{Synthesis Results, QUAP of FIR Filter for 3 Different Implemeted Cases. Power Reduction Is Measured with Respect to Case 1.}
\label{tbl_fir_result}
\centering
\begin{tabular}{|c|c|c|c|c|c|c|}
\hline
\multirow{2}{*}{Case} & SNR$_{out}$ & Clock & Area & Power & Power &  QUAP \\
& (dB) & Period (ns) & ($\mu m^2$) & (mW) & Reduction (\%) & $\div10^4$ \\
\hline
WL = 16, & \multirow{2}{*}{25.35} & \multirow{2}{*}{4.78} & \multirow{2}{*}{1.22\e{5}} & \multirow{2}{*}{3.63} & \multirow{2}{*}{N.A.} & \multirow{2}{*}{N.A.}\\
VBL = 0 & & & & & &\\
\hline
WL = 16, & \multirow{2}{*}{25.0} & \multirow{2}{*}{4.78} & \multirow{2}{*}{1.07\e{5}} & \multirow{2}{*}{3.01} & \multirow{2}{*}{17.1} & \multirow{2}{*}{13.1}\\
VBL = 13 & & & & & &\\
\hline
WL = 14, & \multirow{2}{*}{23.1} & \multirow{2}{*}{4.78} & \multirow{2}{*}{1.13\e{5}} & \multirow{2}{*}{2.91} & \multirow{2}{*}{19.8} & \multirow{2}{*}{7.73}\\
VBL = 0 & & & & & &\\
\hline
\end{tabular}
\end{table}

The quality is assumed to be (SNR$_{out}$)$^2$, as did the reference \cite{gupta}. Since in Eq. 3, power saving and area saving are both variables that related to hardware characteristics, SNR$_{out}$ should increase by the power of 2, to give an equal weight to quality. As seen in Table IV, the implementation which utilizes Broken-Booth Multiplier reduces power consumption by 17.1\% at the cost of 0.4dB SNR reduction and comparing to case number 3, it improves QUAP by 70\%.

%%%%%%%%%%%%%%%%%%%%%%%%%%%%%%%%%%%
%					Conclusion
%%%%%%%%%%%%%%%%%%%%%%%%%%%%%%%%%%%
\section{Conclusion}
In this paper an approximate signed Booth multiplier is proposed which saves power consumption form 28\% to 58.6\% and area from 19.7\% to 41.8\% for different word lengths in comparison to a regular Booth multiplier. The Mean Squared Error (MSE) introduced by this approximate multiplier varies with word length and approximation level, from 0.25 to 8.33\e{7}. To compare the proposed approximate multiplier with two previous works in the literature, Verilog models were prepared and synthesized and then compared using the average power-delay product and MSE criteria. The proposed multiplier shows a reasonable and acceptable performance. To demonstrate an application for the proposed multiplier, an FIR filter was implemented once utilizing accurate multiplier and again using the proposed multiplier. A previously introduced figure of merit (FOM) was used to compare the performance of these versions of the filter implementation. The filter which is using our multiplier has the best FOM.

%% conference papers do not normally have an appendix
%
%
%% use section* for acknowledgement
%\section*{Acknowledgment}



\bibliographystyle{IEEEtran}
\bibliography{cads}

\begin{thebibliography}{10}
\providecommand{\url}[1]{#1}
\csname url@samestyle\endcsname
\providecommand{\newblock}{\relax}
\providecommand{\bibinfo}[2]{#2}
\providecommand{\BIBentrySTDinterwordspacing}{\spaceskip=0pt\relax}
\providecommand{\BIBentryALTinterwordstretchfactor}{4}
\providecommand{\BIBentryALTinterwordspacing}{\spaceskip=\fontdimen2\font plus
\BIBentryALTinterwordstretchfactor\fontdimen3\font minus
  \fontdimen4\font\relax}
\providecommand{\BIBforeignlanguage}[2]{{%
\expandafter\ifx\csname l@#1\endcsname\relax
\typeout{** WARNING: IEEEtran.bst: No hyphenation pattern has been}%
\typeout{** loaded for the language `#1'. Using the pattern for}%
\typeout{** the default language instead.}%
\else
\language=\csname l@#1\endcsname
\fi
#2}}
\providecommand{\BIBdecl}{\relax}
\BIBdecl

\bibitem{mahdiani}
H.~R. Mahdiani, A.~Ahmadi, S.~M. Fakhraie, and C.~Lucas, ``Bio-inspired
  imprecise computational blocks for efficient vlsi implementation of
  soft-computing applications,'' \emph{IEEE Trans. Circuits Syst. I}, vol.~57,
  no.~4, pp. 850--862, 2010.

\bibitem{venka}
R.~Venkatesan, A.~Agarwal, K.~Roy, and A.~Raghunathan, ``{MACACO}: Modeling and
  analysis of circuits for approximate computing,'' in \emph{Proc. CAD}, 2011,
  pp. 667--673.

\bibitem{kulk}
P.~Kulkarni, P.~Gupta, and M.~Ercegovac, ``Trading accuracy for power with an
  underdesigned multiplier architecture,'' in \emph{Proc. VLSI Design}, 2011,
  pp. 346--351.

\bibitem{kelly}
D.~R. Kelly, B.~J. Phillips, and S.~F.~K. Al-Sarawi, ``Approximate signed
  binary integer multipliers for arithmetic data value speculation,'' in
  \emph{Proc. DASIP}, 2009, pp. 97--104.

\bibitem{yin}
K.~Khaing~Yin, G.~Wang~Ling, and Y.~Kiat~Seng, ``Low-power high-speed
  multiplier for error-tolerant application,'' in \emph{Proc. EDSSC}, 2010, pp.
  1--4.

\bibitem{ning}
Z.~Ning, G.~Wang~Ling, Z.~Weija, Y.~Kiat~Seng, and K.~Zhi~Hui, ``Design of
  low-power high-speed truncation-error-tolerant adder and its application in
  digital signal processing,'' \emph{IEEE Trans. VLSI Syst.}, vol.~18, no.~8,
  pp. 1225--1229, 2010.

\bibitem{gupta}
V.~Gupta, D.~Mohapatra, S.~P. Park, A.~Raghunathan, and K.~Roy, ``{IMPACT}:
  Imprecise adders for low-power approximate computing,'' in \emph{Proc.
  ISLPED}, 2011, pp. 409--414.

\bibitem{sampson}
A.~Sampson, W.~Dietl, E.~Fotuna, D.~Gnanapragasam, L.~Ceze, and D.~Grossman,
  ``Ener{J}: Approximate data types for safe and general low-power
  computation,'' in \emph{Proc. PLDI}, 2011, pp. 164--174.

\bibitem{moha}
D.~Mohapatra, V.~K. Chippa, A.~Raghunathan, and K.~Roy, ``Design of
  voltage-scalable meta-functions for approximate computing,'' in \emph{Proc.
  DATE}, 2011, pp. 1--6.

\bibitem{weste}
N.~H.~E. Weste and D.~M. Harris, \emph{CMOS VLSI Design: A Circuits and Systems
  Perspective}, 4th~ed.\hskip 1em plus 0.5em minus 0.4em\relax Boston, MA:
  Pearson Education, 2011.

\bibitem{oppen}
A.~V. Oppenheim, R.~W. Schafer, and J.~R. Buck, \emph{Discrete-Time Signal
  Processing}.\hskip 1em plus 0.5em minus 0.4em\relax Upper Saddle River:
  Prentice Hall, 1999.

\bibitem{shim}
B.~Shim and N.~R. Shanbhag, ``Energy-efficient soft error-tolerant digital
  signal processing,'' \emph{IEEE Trans. VLSI Syst.}, vol.~14, no.~4, pp.
  336--348, 2006.

\end{thebibliography}
%\begin{thebibliography}{1}
%\end{thebibliography}

\end{document}